\documentstyle[12pt, aasms4]{article}

\begin{document}

\title{HST Photometry of the Fornax dSph Galaxy: cluster 4 and
its field.}

\author{R. Buonanno\altaffilmark{1}, C.E. Corsi\altaffilmark{1}, M.
Castellani\altaffilmark{1}, 
G. Marconi\altaffilmark{1},
F. Fusi Pecci\altaffilmark{2},
R. Zinn\altaffilmark{3}}

\lefthead{Buonanno et al.}
\righthead{HST photometry of cluster 4 in Fornax}

\altaffiltext{1}{Osservatorio Astronomico di Roma, 
Via Frascati 33, I-00040 Monte Porzio Catone, Rome, Italy} 
\altaffiltext{2}{Stazione Astronomica di Cagliari, 
I-09012 Capoterra, Cagliari, Italy}
\altaffiltext{3}{Department of Astronomy, 
Yale University, Box 208101, New Haven, Connecticut 06511}		

% buonanno@coma.mporzio.astro.it 
% zinn@astro.yale.edu

\date{Received; accepted}

\begin{abstract}

Using observations from the {\it Hubble Space Telescope} archive,
color-magnitude diagrams (CMDs) have been constructed for globular
cluster 4 in the Fornax dSph galaxy and its surrounding field.  These
diagrams extend below the main-sequence turnoffs and have yielded
measurements of the ages of the populations.

The most prominent features of the CMD of the Fornax field population
are a heavily populated red clump of horizontal branch (HB) stars, a
broad red giant branch (RGB), and a main sequence that spans a large
range in luminosity.  In this CMD, there are very few stars at the
positions of the HBs of the five globular clusters in Fornax, which
suggests that only a very small fraction of the field population
resembles the clusters in age and chemical composition.  The large
span in luminosity of the main-sequence suggests that star formation
began in the field $\simeq$12 Gyrs ago and continued to $\simeq$0.5
Gyr ago.  There are separate subgiant branches in the CMD, which
indicate that the star formation was not continuous but occurred in
bursts.

The CMD of cluster 4 has a steep RGB, from which we estimate
[Fe/H]$\simeq$--2.0.  This is considerably lower than estimates from
the integrated light of the cluster, and the origins of this
discrepancy are discussed.  Cluster 4 has a very red HB, and is
therefore a prime example of the second parameter effect.  Comparisons
of cluster 4 with the other Fornax clusters and with M68, a very metal
poor globular cluster of the galactic halo, reveal that cluster 4 is
$\simeq$3 Gyrs younger than these other clusters which have much bluer
HBs.  This age difference is consistent with the prediction that age
is the 2$^{nd}$ parameter to within the uncertainties.  

The CMD of cluster 4 is virtually identical to that of the unusual
globular cluster of the galactic halo, Ruprecht 106, which suggests
that they have very similar ages and chemical compositions.  We discuss
the
possibility that cluster 4 also resembles R106 in having a larger [Fe/H]
than is indicated by its steep RGB and also a lower [$\alpha$/Fe]
ratio than is usual for a globular cluster, as indicated by some recent
observations of R106.

The CMDs of the five Fornax clusters indicate that cluster age is a
major but probably not the sole 2$^{nd}$ parameter.  Buonanno {\it et
al.} (1998a) concluded that cluster density probably influenced
the HB morphologies of clusters 1, 2, 3, and 5.  Despite a
very large difference in central density, the HBs of cluster 4 and
R106 are very similar.  This suggests that density may act as a
2$^{nd}$ parameter in clusters which have HBs on the verge
of moving towards the blue or
already blue for another
reason, such as very old age.

\end{abstract}


\pagebreak

%%%%%%%%%%%%%%%%%%%%%%%%%%%%%%%%%%%%%%%%%%%%%%%

\section{Introduction}
The dwarf spheroidal (dSph) galaxies of the Local Group provide an
opportunity to study star by star the histories of star and star
cluster formation in galaxies of the very lowest mass.  While once
these galaxies were thought to be relatively simple systems composed
entirely of very old stars, they are now known to have experienced
much more complex histories (see Da Costa 1998 and Mateo 1998 for
recent reviews).  The Fornax dSph galaxy is no exception, for contains
very old globular clusters (Buonanno {\it et al.} 1998a, hereafter
BEA98), many stars of intermediate age, and even stars younger than
0.1 Gyr (Stetson {\it et al.} 1998).  There remain, however, many unanswered
questions regarding Fornax, which is one of the most thoroughly
studied galaxies of this type.  For example, did the star formation in
Fornax occur in bursts, as it did in the Carina dSph (Smecker-Hane
{\it et al.} 1994; Smecker-Hane {\it et al.} 1996; Mighell 1997), or was 
it relatively continuous? 
Also, what fraction of the field population in Fornax resembles its 5 globular
clusters in age and chemical composition?

For more than two decades there has been speculation (see Zinn 1993 and
Mateo 1998 for reviews) that the outer halo of the Milky Way was
created by the tidal destruction of dwarf galaxies, particularly ones
that resembled the dSph galaxies.  Although the Sagittarius dSph
galaxy is now in the process of being destroyed and blended into the
halo, it remains to be seen whether or not this was the only or even
the primary mechanism by which the halo formed.  Additional
observations of the cluster and field populations in Fornax and the
other dSph galaxies are needed to test this idea, which is not so
simple because the dSph galaxies that have survived to the present may
have undergone much more evolution than the hypothetical ones that
were destroyed in the past, perhaps at widely different epochs.

The most massive dSph galaxies, Fornax and Sagittarius, have their own
systems of globular star clusters, which are very interesting objects
in their own right.  In the case of Fornax, the clusters can be
considered to be at essentially the same distance from us, which means
that comparisons among them are independent of the uncertain
distance scale.

The several previous investigations of the Fornax clusters (e.g.,
Buonanno {\it et al.} 1985, hereafter BEA85; Beauchamp {\it et al.}
1995; Smith {\it et al.} 1996, 1997, 1998; BEA98) have shown that they
differ in many important characteristics, which make them particularly
valuable for studying stellar evolution and the possible connection
between it and cluster dynamics.  They present clear evidence of the
``second parameter effect'' (e.g, Lee, Demarque \& Zinn 1994;
Sarajedini, Chaboyer \& Demarque 1997), because the differences among
the horizontal branch (HB) morphologies of clusters 1, 3 and 5 (the
most metal-poor of the Fornax clusters) are not explained by their
small differences in metallicity.  The luminosity profiles of clusters
1 and 2 resemble those of typical Galactic globular clusters with
large core radii and truncated halos, while clusters 3, 4 and 5 have
smaller core radii and extended halos (Webbink 1985).  The range of
central densities spans from log $\rho$$_0$=0.454 for cluster 1 to
3.83 and 3.93 for clusters 3 and 4 respectively (Webbink 1985).
According to previous measurements, there is a substantial range in
metallicity among the clusters, from [Fe/H]=--2.2 for clusters 1 and 5
(BEA98) to [Fe/H]=--1.40 for cluster 4 (Beauchamp {\it et al.} 1995).  Given
the small back to front range in distance modulus of Fornax and this
wide range in [Fe/H], BEA85 noted that the Fornax clusters may provide
a precise measurement of the dependence of HB luminosity on [Fe/H].

Recently, BEA98 used the WFPC2 of the Hubble Space Telescope to
construct the color-magnitude diagrams (CMDs) of clusters 1, 2, 3 and
5, in order to measure their ages and to explore the connection
between HB morphology and cluster density. They concluded that the
four clusters have the same age to within 1 Gyr and that this age
difference is too small to explain the observed differences in HB
morphology unless the HB is more sensitive to age than previously
thought. In addition, they noted that a correlation exists between the
HB morphologies and the central densities of the clusters.
Unfortunately, BEA98 could not obtain an estimate of the slope
M$_V$(HB) {\it vs} [Fe/H] relationship, because the range in metal
abundance among these clusters is too small
([Fe/H]$_{cl.2}$--[Fe/H]$_{cl.1}$ $\simeq$0.4$\pm$0.3).
  
In this paper we present the results of a study of the CMD of Fornax
cluster 4 and its surrounding field.  As a continuation of our
previous work on the other Fornax clusters, we intended to focus
primarily on the age of the cluster 4 and the M$_V$(HB)--[Fe/H]
relation, but our results have motivated a shift in emphasis which is
explained below.  Previous investigations of the CMD of cluster 4
(BEA85; Beauchamp {\it et al.}  1995) have been
hampered by the high density of stars in the cluster and in the
surrounding field, which is near the center of the galaxy.  Because
the images from Hubble Space Telescope (HST) have much higher
resolution than the previous ones from ground-based telescopes, they
have enabled us to do photometry to below the main-sequence turnoff in
cluster 4.

%%%%%%%%%%%%%%%%%%%%%%%%%%%%%%%%%%%%%%%%%%%%%

\section{Observations and reductions}

The observations consisted of two 1100 {\it sec} and one 200 {\it sec}
exposures in each of the two filters F555W and F814W of the
refurbished WFPC2.  The data were retrieved electronically from
ESO/ST-ECF archive (Proposal n. GTO/WFC 5637, P.I.: Westphal, data taken 
in March 1995).
Cluster 4 is located within the area of WF3, while the areas of WF1
and WF2 sample essentially only the field population of the dSph galaxy.
Fig. 1 shows the observed field with a few isolated stars marked for
the purpose of identification.

The photometry of the stars in the three WFs (scale 0.0996 {\it
arcsec/pixel}) were performed with DAOPHOT II using the hybrid
weighted technique described by Cool and King (1995).  To push the
detection of stars to the faintest possible limit, we coadded the deep
images taken with the same filter.  We first performed the DAOPHOT
(Stetson 1987) detection step, did photometry of the detected stars,
and then subtracted their images from the frame.  We then performed
the detection procedure a second time, which created a second list of
stars to be added to the first.  This procedure was very effective at
detecting stars, even those hidden within the outskirts of the PSFs of
the brighter stars.  Because PC field is relatively small in size, it
is unlikely to add any new information about the stellar populations
of Fornax, and for this reason, we did not include it in our
photometric reductions.

For each chip and each filter, the PSF was built using at least 15
bright and isolated stars in each frame.  Corrections to 0.5''
aperture were made in each case, and the F814W and F555W instrumental
magnitudes were transformed into the WFPC2 ``ground system'' using
Eq. 6 of Holtzman {\it et al.} (1995).

Before starting the reductions, the long-exposure were processed by
routines of ROMAFOT expressely developed to eliminate the cosmic rays
by processing a series of frames with a median filter.  Photometry was
also performed using ROMAFOT in order to compare it with the technique of
Cool and King (1995).  The more complex but slower routines of ROMAFOT, which
were developed for crowded fields, yielded fully compatible results.
ROMAFOT was not used for our final reductions because the Fornax
fields are relatively uncrowded.

%%%%%%%%%%%%%%%%%%%%%%%%%%%%%%%%%%%%%%%%%%%%%%

\section{Color-Magnitude diagrams}

Fig. 2 shows the V {\it vs} (V--I) CMD for the three WFPC2 fields.
On the left of the figure, the photometric errors at the different
magnitude levels are reported.  Local position and photometry of the
brightest stars are reported in Table 1a; the photometry for all
the stars measured in this paper, with coordinates referred to the
center of the cluster, can be found at {\tt
http://www.mporzio.astro.it/$\sim$mkast/data.html}, while 
the same photometry, referred to the bottom-left
corner of each chip, is reported in Tables 1b, 1c, 1d respectively
for wide field cameras WF2, WF3 and WF4 (note: such tables will be available
on the electronic version of this paper, when published).

The major features to note in Fig. 2 are, first, the bifurcated Red
Giant Branch (RGB) at V$\leq$19.5, which indicates the presence of two
stellar populations of different metallicity, second, the two distinct
horizontal branches (HBs), reminiscent of an ``old'' HB located at
V$\simeq$21.5, and an intermediate-age red HB clump which is about
0.25 mag brighter and 0.1 mag redder than the old HB and, third, the
young main sequence (MS) on the blue side of the diagram.  Each of
these features clearly indicates that the CMD of Fig. 2 contains at
least two distinct populations, which most likely belong to cluster 4
and to the field of Fornax.

To investigate this further, we first plotted in Fig. 3 the stellar
density profile in units of stars per square arcmin, with the origin at
the center of cluster 4.  We then selected two samples that we believe
are more representative of the cluster and of the field
populations. The {\it cluster sample} consists of stars that lie at a
distance R$\leq$18 arcsec from the cluster center, i.e. the region
where the mean star density is at least twice the field density.  The
{\it field sample} is drawn from the region where the density of stars is
constant, which is at R$\geq$ 60 arcsec.

%%%%%%%%%%%%%%%%%%%%%%%%%%%%%%%%%%%%%%%%%%%%%%%%%%%

\subsection{The field of Fornax around cluster 4} 

The V {\it vs} (V--I) CMD of the sample of the field of Fornax, as
defined above, is displayed in Fig. 4.

This diagram, which is based on 4742 stars, clearly shows a prominent
population of relatively blue stars and a rich clump of stars centered
at (V--I)$\simeq$0.97. Both features are characteristic of an
intermediate-age population, whose presence in Fornax has been
suggested by several other workers (e.g., BEA85; Beauchamp et
al. 1995; Stetson {\it et al.} 1998; Da Costa 1998).  Another important feature in
Fig. 4 is the RGB extending to V=18.63 and (V--I)=1.66 which,
however, is broader than expected on the basis of the photometric
errors alone.  This could be caused by a spread in metallicity of the
Fornax field stars and/or the presence of a few asymptotic giant
branch stars (AGB) around V$\simeq$20.5.  The more extensive but less
deep CMDs of the field of Fornax that have been published previously
have indicated that a significant range in [Fe/H] exists (see Da
Costa's 1998 review).  Two very red stars have not
been plotted in Fig. 4.  They lie at V=18.23, V--I=2.64 and V=19.54,
V--I=3.06; for both colors and magnitudes such stars could be AGB stars,
or also BSs progeny similar to that found in 47 Tuc (Montegriffo {\it et al.}
1997).

The fiducial line of the RGB of the Fornax field, as
defined by our photometry, is listed in Table 2.
The procedure adopted for determine the ridge-line discarding the outliers
is based on the construction of a series of histograms in CMD boxes.
The boxes have dimensions $\pm$ 0.1 mag in V,
$\pm$ 3 times the photometric error in V--I evaluated at each magnitude level.
After such a selection, the resulting "cleaned" CMD has been checked by eye.
We adopted the mode of the histogram in each bin as
representative of the true mean color at a given magnitude and the resulting
curve has been smoothed.
The errors in color listed in column 3 of Table 2 are the standard error
calculated within each box.

As noted above, HB of the field population consists primarily of a red
clump of stars.  The few bluer stars of roughly the same magnitude may
belong to a relatively sparsely populated second HB, which is much
more evident in the CMD published by Stetson {\it et al.} 
(1998, see also Da Costa
1998).  To measure the luminosity of the red clump in our diagram, we
selected the stars in Fig. 4 between 20$\leq$V$\leq$22 and
0.88$\leq$(V--I)$\leq$1.04 and found a mean value
V$_{clump}$=21.25$\pm$0.20, where the error is the standard error of
the mean.  Subsequently, we found for the RGB color at the HB level
$(V-I)_{g,field}\simeq1.04\pm$0.05, where the uncertainty is the
combination of the uncertainties in V$_{clump}$ and the ridge line
of the RGB.

To estimate the average metallicity of the field population we used
the parameter {\it ``sl''} defined by Buonanno {\it et al.}
(1993) as {\it
sl}=$(V-I)_{-2.4}-(V-I)_{-1}$, where $(V-I)_{-i}$ is the color of the
RGB at {\it i} magnitudes brighter than the HB.  The parameter {\it
``sl''} has been calibrated using six template clusters (see Table 5 of
Buonanno {\it et al.} 1993).

From the data in Table 2, we find {\it sl} = $1.51-1.19=0.32$ and,
then, $[Fe/H]=-1.36\pm0.16$, where the uncertainty in [Fe/H] is the
combination of the assumed uncertainty in the metallicity of
calibrators ($\sigma_{[Fe/H]}$=0.15) and in the estimate of the RGB
average color ($\sigma$$_{(V-I)}$=0.05).  Adopting the calibration of
Sarajedini (1994), E(V--I)=(V--I)$_g$--0.1034[Fe/H]--1.100, we obtain
E(V--I)=0.08$\pm$0.05 for the reddening of Fornax.  These measurements
of mean [Fe/H] and E(V--I) are in good agreement with previous ones
(e.g., BEA85; Beauchamp {\it et al.} 1995).
 
The photometry in Fig. 4 can be used to establish an upper limit to
the true distance modulus of the Fornax dSph.  We start by
assuming that the bright star at V=18.63 and (V--I)=1.66 is actually
at the tip of the RGB (TRGB).  According to Lee {\it et al.} (1993), the I
magnitude of the TRGB is a weak function of metallicity and therefore
is a distance indicator.  Adopting M$_I$=--4.0$\pm$0.1 as
the absolute I magnitude of the TRGB, the apparent distance modulus
of Fornax turns out to be (m--M)$_I$=16.97+4.0=20.97$\pm$0.10.  Then
with
A$_I$=1.29 E(V--I) and A$_V$=2.66 E(V--I) (Cardelli {\it et al.}
1989), we obtain for the true distance modulus
(m--M)$_0$=(m--M)$_I$--A$_I$=20.97--0.10=20.87$\pm$0.11, and
(m--M)$_V$=20.87+0.21=21.08$\pm$0.11.  Because the TRGB has been
defined by a single star, this distance modulus must be regarded as
an upper limit.

A more definite estimate of the distance modulus can be
obtained from Fig. 5, where we have superimposed the ridge lines of
the globular cluster M5 (Johnson \& Bolte 1998) and the Fornax field.
The metallicity of M5 is [Fe/H]=--1.40$\pm$0.06 (Zinn \& West 1984),
which is essentially the same as that found here for mean [Fe/H] of
Fornax.  In Fig. 5 the M5 loci were shifted in color by E(V--I)=0.04,
to account for the reddening of Fornax (E(V--I)$_{M5}$=0.04, Zinn \&
West 1984), and vertically by
$\Delta$V=6.55.

To obtain the distance modulus of M5 and therefore that of Fornax from
the match of the RGBs in Fig. 5, we use the luminosity of the HB in
M5, V(HB)=15.11$\pm$0.1 (Buonanno {\it et al.} 1988), to set the distance
scale.  Adopting M$_V(RR)$=0.82+0.17[Fe/H] from Lee {\it et al.} (1990),
we obtain $M_V(RR)=0.58$ and, then, (m--M)$_V^{M5}$=$14.53\pm 0.10$.
Consequently, we obtain
(m--M)$_V^{Fornax}$=14.53+6.55--2.66(0.04)=20.97$\pm$0.10, which is
consistent with the upper limit from TRGB.  The true distance modulus
by this method is 20.76$\pm$0.10.  It is important to note that this
procedure is relatively insensitive to age differences between the
population of Fornax and that of M5, because the age sensitivity of
the RGB is small, being $\Delta$(V--I)/$\Delta$t$\simeq$0.007 mag/Gyr
(Da Costa \& Armandroff 1990).

Another estimate of the distance modulus of Fornax can be obtained
from the four globular clusters that are similar in age to the
globular clusters in the Milky Way.  Using the values of V(HB),
E(V--I), and [Fe/H] that BEA98 list in their table 2 for clusters
1, 2, 3, and 5, and the same relation for M$_V(RR)$ as above, we
obtain an average (m--M)$_0$=20.62$\pm$0.08.  This value
is the same, to within the errors, as our estimate from the RGB of the
Fornax field, which builds confidence in our photometry.  

Given the present considerable uncertainty over the distance scales
for RR Lyrae variables and globular clusters (cf. Chaboyer {\it et
al.} 1998, Popowski \& Gould 1998), the major uncertainty in the
distance modulus of Fornax is the scale applied to parameters such as
V(HB) and not their observational errors.  Since the scale of Lee {\it
et al.} (1990) lies near the middle of the range of several
alternatives, we suggest the value of 20.68$\pm$0.20 for the true
distance modulus of Fornax.

%%%%%%%%%%%%%%%%%%%%%%%%%

\subsection{Star formation history of Fornax}

The large range in luminosity of the main-sequence of the Fornax field
population (see Fig. 4), indicates that Fornax has experienced a long
history of star formation.  While this has been detected previously by
several teams of investigators (e.g., BEA85; Beauchamp {\it et al.}
1995;
Stetson 1997; Da Costa 1998), our data provide some additional
information.

The bright limit of the main-sequence in Fig. 4 is V$\simeq$21.0,
which corresponds to M$_V$$\simeq$+0.1.  With the assumption that these
stars are near the stage of central exhaustion of core H burning, the
isochrones of Bertelli {\it et al.} (1994) indicate an age of about
0.5 Gyrs, and this result depends only weakly on the assumed metal
abundance (see Fig. 9 in Bertelli {\it et al.} 1994).  While this result is
very remarkable for a galaxy which was once thought consist of a
single population of very old stars, the CMD presented by Stetson
(1997) and also discussed by Da Costa (1998) indicates that Fornax
contains a few stars younger than 0.1 Gyr.

At fainter magnitudes in Fig. 4, one sees a subgiant branch (SGB)
peeling off from the main-sequence.  Only in stellar populations older
than about 3 Gyrs is the Hertzsprung Gap closed by the development of
a well-populated SGB (see for example the discussion of Hardy {\it et
al.} 1984 on the LMC bar).  With reasonable estimates for the metal
abundance of Fornax, the isochrones of Yale (Demarque {\it et al.} 1996)
shown in Fig. 6 and that of Chieffi, Straniero
\& Limongi (private communication) in Fig. 7 indicate an age of about 2-4
Gyrs for the population responsible for the brightest SGB in Fig. 4.
The core He burning phase of this population will produce a red clump
centered near M$_V$=0.3 (Bertelli {\it et al.} 1994).  This corresponds to
V=21.2 in Fornax and coincides with the red clump in Fig. 4.  The
fainter SGBs in Fig. 4 indicate the presence of older stars, which
have core He burning phases that slowly decrease in M$_V$ with the
increasing age of the population.
Thus, the prominence of the red clump in Fornax can be
attributed to the funneling of stars of wide range of ages to
approximately the same point in the CMD.  The populations that are
younger than 3 Gyrs also have core He burning phases, which produce
brighter but shorter lived red clumps.  The dispersion in magnitude of
the red clump in Fornax is probably do the presence of a few of these
stars in addition to the evolution of older stars from the red clump
to the AGB.

As noted above one of the striking features of Fig. 4 is the absence
of an HB resembling the ones seen in old globular clusters.  The 4
oldest clusters in Fornax, clusters 1, 2, 3, and 5, have HBs that
extend over wide ranges in color and include many RR Lyrae variables
(see BEA98 and references therein).  There are at most a very few
stars in Fig. 4 that can be attributed to such an HB population or
even to a red HB resembling the one in cluster 4, which is shown below
to be about 3 Gyrs younger than these other clusters.  Since a weakly
populated blue HB is seen in the photometry of a larger field (Stetson
{\it et al.} 1998, see also Da Costa 1998), and since Mateo (1998)
reports that more than 400 RR Lyrae variables have been discovered in
Fornax, the field population of Fornax does have an old, metal-poor
component.  Evidently, it is a minor one in comparison to the
intermediate-age populations.  Therefore the absence of BHB stars in
the CMD of the field (r$\ge$60 arcsec) could be explained with the
expected number (about 1) of RR-Lyrae variables obtained scaling the
Stetson (1998) {\it et al.} results to the small area covered by our
data.

Finally, in contrast to the CMD constructed by Stetson {\it et al.}
(1998) which provided no evidence for bursts of star formation like
the ones that are so evident in the CMD of the Carina dSph
(Smecker-Hane {\it et al.} 1994; 1996), our deeper and probably more
precise CMD from HST observations reveals signs of a variable star
formation rate.  If star formation was continuous in our field, we
would expect to see a smooth distribution of stars between the
main-sequence and the RGB in Fig. 4.  Instead, there appear to be gaps
between separate SGBs, which are suggestive of separate bursts of star
formation.

This is illustrated in Fig. 6, which is an enlarged version of Fig. 4.
Using our best estimates for (m--M)$_V$ (20.89) and E(V--I) (0.08), we
have also plotted in Fig. 6 the ridge line of M5 (from Johnson \&
Bolte 1998) and the Yale isochrones for ages of 1, 2, 4, and 7 Gyrs
(Demarque {\it et al.} 1996).  This diagram shows that while the ridge
line of M5 matches the giant branch, as was illustrated previously in
Fig. 5, its SGB forms a lower limit on the luminosities of the
subgiants in this field of Fornax.  Under the assumption of similar
chemical compositions, which is reasonable given the coincidence of
the giant branches, this suggests that the majority of the Fornax stars
are younger than M5.  The near absence of any stars resembling the HB
stars in M5 is also consistent with this.  To be precise about the age
distribution of the Fornax stars requires much more information about
their chemical compositions than we currently have.  Consequently, the
following estimates from the comparison with the isochrones should be
considered only very rough estimates. 

Since there are few constraints on the metal enrichment history of
Fornax, we have chosen to plot isochrones of different ages but for
the same chemical composition (Y=0.23, Z=0.004, and a solar mix of
elements).  Salaris {\it et al.} (1993) have shown that in the region
of the main-sequence turnoff (TO) and the SGB the isochrones for the
solar mixture closely approximate ones for a mixture where the
$\alpha$ elements are enhanced.  According to their relationship, the
isochrones used here are appropriate for a $\alpha$-enhancement of a
factor of 3 and metal content of Z=0.0018 ([Fe/H]$\simeq$--1.05), which
is only slightly more metal rich than some measurements for M5 
([Fe/H]$\simeq$--1.11,
Carretta \& Gratton 1997).  The main-sequence and subgiant region of
M5 are approximated by a 12 Gyr isochrone of this composition.  The
younger stellar populations of Fornax may have more nearly solar
ratios of [$\alpha$/Fe], as do populations of similar age in the Milky
Way.  This change in [$\alpha$/Fe] is thought to be caused by an
increase in the abundances of the Fe peak elements once type Ia
supernovae begin to explode.  Hence, these isochrones for a solar mix
and Z=0.004 ([Fe/H]$\simeq$--0.7) may not be far off for these
populations, although we emphasize that there currently no
observational evidence to support this.

The comparison of the isochrones in Figs. 6 and 7 suggest that this field of
Fornax had major episodes of star formation at ages of roughly 7, 4,
and 2.5 Gyrs, with relatively little star formation in between.  We
must emphasize that these ages are sensitive to the choices for the
distance modulus and the reddening of Fornax and the distance scale,
as well as the chemical compositions of the isochrones.  The most
important point is that a highly variable star formation rate, as
illustrated by these separate isochrones, is consistent with the
observed SGBs.  Also, the luminosity and the color of the
main-sequence is consistent with star formation continuing until
$\leq$ 1 Gyr ago.

Note that these conclusions are independent of the particular  set
of isochrones adopted. This is clearly illustrated in Fig. 7 where another
set of
isochrones (Chieffi, Straniero \& Limongi, private communications) 
for the same metallicities and
ages are plotted with the same data that was plotted in Fig. 6.
One addictional interesting feature of Fig. 7 is the luminosity of the core
helium burning stars which coincides with the observed clump and
supports the adopted distance modulus.

%%%%%%%%%%%%%%%%%%%%%%%%%%%%%%%%%%%%%%%%%%%%%%%%%%%%%%%%%%%%%%%%%

\subsection{Cluster 4}

The V {\it vs} V--I CMD of cluster 4 is displayed in Fig. 8. The
diagram is based on 1343 stars within the distance of 18 arcsec from
the cluster center, and reaches V$\simeq$26.0. The overall morphology
is similar to that of a Galactic globular cluster with a well
developed RGB extending to V$\simeq$18.28 and (V--I)$\simeq$1.63.  The
subgiant and the TO regions, although clearly delineated, appear
somewhat contaminated by the Fornax field.

The V magnitude of the HB was determined by an iterative procedure which
rejected the 2$\sigma$ outliers from the mean value.  It yielded
V$_{HB}$=21.52$\pm$0.05.
The fiducial line of the CMD of Fornax cluster 4 are listed in Table 3.
In order to derive the fiducial of the cluster we followed the same procedure 
adopted for the field (see section 3.1 for details and error estimates). 
In the TO region the determination of the fiducial could be affected by the
presence of field stars along the MS and in particular by the bright and
blue stars lying above the SGB. We carefully checked that such stars have 
been rejected as outliers by our selection procedure. Moreover having adopted 
the mode of the histogram, the presence of field stars
should not be a major problem if one accept that the large majority
of the stars belong to the cluster.

\subsection{The metallicity and reddening of cluster 4}

The RGB of cluster 4 is the bluer of the two branches seen in Fig. 2,
which suggests that it is more metal poor than the mean abundance of
the field population ([Fe/H]=--1.36, see above).  To estimate the
metallicity of cluster 4, we followed the same procedure that we used
for the field and computed from Table 3 the parameter $sl =
(V-I)_{-2.4}-(V-I)_{-1}=1.305-1.103=0.202$.  From inspection of Table
5 of Buonanno {\it et al.} (1993) we conclude that the metallicity of
cluster 4 is intermediate between M15 and NGC6397, and from linear
interpolation we obtain [Fe/H]=--2.01$\pm$0.14, a metallicity very
similar to those of the other Fornax clusters.  Once the metal
abundance is known, the reddening can be estimated from the color of
the RGB.  Adopting [Fe/H]=$-2.01\pm0.20$ and from our measurement of
$(V-I)_g=1.028\pm0.05$, we obtain E(V--I)=0.14$\pm0.05$ for the
reddening of cluster 4.

Since nearly all previous determinations found much higher values of
[Fe/H] ($\simeq-1.3$ see below), we have also estimated the
metallicity and the reddening of cluster 4, by placing the ridge line
of its RGB in the M$_I$, (V--I)$_0$ plane following Da Costa \&
Armandroff (1990). Although this procedure is not independent of the
method applied above, it checks whether or not the whole RGB is
consistent with the value of [Fe/H] that was inferred from $sl$.
Fig. 9 shows that for E(V--I)=0.15 and (m---M)$_I$=20.9 the RGB of
cluster 4 lies between those of M15 and NGC6397, confirming that
[Fe/H]$_{cl4}$$\simeq$--2.00 (note that neither V$_{HB}$ nor
(V--I)$_g$ was used in this comparison).  On the basis of the previous
measurements, one would expect the RGB of cluster 4 to match that of
NGC 1851 ([Fe/H]=--1.29), but as Fig. 9 shows this is totally
inconsistent with the slope of the RGB of cluster 4.  By adjusting the
distance modulus and reddening of cluster 4 within acceptable limits,
one can force an approximate match of its RGB to that of M2
([Fe/H]=--1.58).  For the following reasons, we believe cluster 4 is
is more metal-poor than this.

In Fig. 10, cluster 4 is compared with Fornax cluster 2, which is the
most metal rich of the other Fornax clusters according to several
measurements.  The slope of its RGB indicates [Fe/H]=$-1.78\pm0.20$
(BEA 98), which is within the combined errors of the value obtained
above for cluster 4 by the same technique.  In the lower panel of
Fig. 10 the ridge lines of the two clusters have been plotted after
making the same reddening and extinction corrections for each cluster.
These corrections are based on the reddening that BEA 98 measured for
cluster 2, which is only 0.01 mag larger than the value obtained above
for the reddening of the field near cluster 4.  One can see from this
comparison that although the giant branches of the two clusters run
roughly parallel, the giant branch of cluster 4 is definitely redder.
The HB of cluster 4 is also offset $\simeq0.17$ mag fainter than HB of
cluster 2, which is too large to be due to either a difference in
distance modulus or a modest difference in [Fe/H].  These offsets can
be partially explained by larger reddening for cluster 4, which is
illustrated in the upper panel of Fig. 10 where the reddening of
cluster 4 has been raised to E(V--I)=0.12.  Now the RGBs of the two
clusters match, but the difference in $V_{HB}$ is still surprisingly
large, $\simeq0.09$ mag, if the clusters have the same chemical
composition.  While this match with cluster 2 is acceptable, the
offset in $V_{HB}$ suggests that cluster 4 may be even more metal
poor and somewhat more heavily reddened, as indicated by Fig. 9 and
the $sl$ parameter.

A confirmation of this conclusion is provided by the comparison of the
ridge lines of cluster 4 with those of the metal poor galactic
globular cluster M68 ([Fe/H]=--2.09, Zinn \& West 1984), which is
shown in Fig. 11.  Cluster 4 has been dereddened by 0.15 mag and
shifted by $\Delta$V=0.40 mag to account for the absorption.  The M68
ridge line from Walker (1994) has been dereddened by E(V--I)=0.09 (see
BEA98) and shifted by $\Delta$V=5.50 mag in V to match the HB
luminosity of cluster 4.  Fig. 11 shows that the RGB of cluster 4 is
very similar to that of M68 and confirms the low metallicity estimate
obtained above.  Note, however, that the TO of cluster 4 is
significatively brighter ($\Delta V=0.2\pm0.1$ mag) than the TO of
M68. Adopting $\Delta V = 0.07 $ mag/Gyr (Vandenberg, Stetson \& Bolte
1996), we estimate that cluster 4 is approximatively $2.9\pm1.5$ Gyr
younger than M68.  According to the upper panel of Fig. 10, the
turnoff of cluster 4 is also brighter than the turnoff in cluster
2.  A more detailed comparison of the ages of cluster 4 and the other
Fornax clusters is made in section 3.5.

The low value of [Fe/H] obtained above disagrees with estimates from 
photometry and spectroscopy of cluster 4's integrated light from blue to 
near infrared wavelengths, which have consistently yielded values of 
[Fe/H] near that of the field population of Fornax, i.e., $\simeq-1.3$
(Harris \& Canterna 1977, Zinn \& Persson 1981, Dubath {\it et al.}
1992, Beauchamp {\it et al.} 1995).  Our value agrees, however, with
the value that Beauchamp {\it et al.} (1995) estimated from their CMD
of cluster 4, which was constructed from ground-based photometry that
barely reached the HB.  Beauchamp {\it et al.} (1995) considered the
inconsistency between the position of the RGB in their CMD and the
conclusions drawn from integrated light measurements a mystery to be
resolved by HST observations.  We believe the much improved CMD
produced by the HST observations has at least partially done this by
conclusively showing that cluster 4 the has a steeply sloped RGB of a
very metal poor cluster.

We have thought of three factors that either individually or, more
likely, collectively may account for the discrepancy with the
inferences from the integrated-light measurements.  These observations
measured either the broad-band colors of the cluster or the strengths
of metal absorption lines, either spectroscopically or
photometrically.  For old globular clusters, there are tight
relationships between these quantities and [Fe/H] because of the
dependence of the color of the RGB on [Fe/H] (e.g., Zinn \& West
1984).  The integrated-light observations have demonstrated that
cluster 4 is redder and has stronger absorption lines than three other
Fornax clusters (2, 3, \& 5) and also other very metal-poor globular
clusters belonging to the Milky Way, such as M68.  The conclusion that
cluster 4 is more metal-rich than these clusters depends critically on
whether or not the relationships established among the globular
clusters of the Milky Way are applicable to it.  The CMD's of Fornax
clusters 2, 3, \& 5 (see BEA98) are similar to that of M68 and other
very metal-poor globular clusters in the Milky Way, and not
surprisingly, there is good agreement between the values of [Fe/H]
that are inferred from their CMD's and from the integrated-light
observations.  However, as shown in Figs. 10, 11, and 12 the CMD of
cluster 4 is much different from those of the other Fornax clusters
and M68 in that it has a much redder HB and a brighter SGB.
Consequently, both the CMD and the integrated-light measurements are
in agreement that cluster 4 is unlike the other Fornax clusters and
M68.  The greater information provided by our CMD suggest that its
relatively red color and stronger absorption lines are not due to a
redder RGB, as has been inferred previously, but to something else,
and a likely candidate is the light contributed by the redder HB and
brighter SGB in cluster 4.

These may not be the only differences, however.  Below we will show
that the CMD of cluster 4 is nearly identical to that of the Milky Way
globular cluster Ruprecht 106 (see Fig. 13), which is younger by about
4 Gyrs than the typical globular cluster in the galactic halo
(Buonanno et al. 1993).  Several studies of the CMD of R106 have shown
that it has a steep RGB that is indicative of [Fe/H]=--1.9 (Buonanno
{\it et al.}  1993; Sarajedini \& Layden 1997).  However,
spectroscopic observations of red giants have yielded significantly
higher values than this (Francois {\it et al.}  1997; Brown {\it et
al.} 1997), and the observations of Brown {\it et al.} (1997) indicate
that R106 has an anomalously low value of the [$\alpha$/Fe] ratio.  A
similar discrepancy exists between the [Fe/H] inferred from the RGB
and spectroscopic measurements for the metal rich young globular
cluster Pal 12, and it too appears to be [$\alpha$/Fe] deficient
compared to other globular clusters (Brown {\it et al.} 1997).  It is
possible that cluster 4 is another example of this phenomenon (see
Fusi Pecci {\it et al.} 1995 and Sarajedini \& Layden 1997 for
discussions of its possible origin).  This would at least partially
explain why the integrated spectrum of cluster 4 has relatively strong
metal lines and yet its RGB is quite steep.

Finally, it is possible that the integrated-light observations have
been contaminated by light from stars belonging to the field
population of Fornax.  Unlike the other globular clusters in Fornax,
cluster 4 lies near the center of the galaxy where the density of the
field population is largest.  Consider, for example, the 3'' $\times$
5.9' slit that Beauchamp {\it et al.} (1995) used to measure the
spectrum of cluster 4.  According to the curve in Fig. 3, which for
this calculation we extrapolated inward, this slit included about 400
stars down to the limit of our photometry.  However, only about 80 of
them (20\%) belong to the cluster.  Because the field population has
a relatively red RGB, mean [Fe/H]=--1.36, and heavily populated red
clump (see above), the removal of only part of the contamination from
the field may produce spuriously red colors and/or metal-line
strengths.

Given that cluster 4 is either very metal-poor like the other Fornax
clusters or (more speculatively) has unusual mix of elements like
R106, it and the other Fornax clusters do not provide a reliable
sample to empirically derive the slope of the M$_V$(HB) vs. [Fe/H]
relationship.  The Fornax clusters do provide, however, important
information on the origin of the second parameter effect.

%%%%%%%%%%%%%%%%%%%%%%%%%%%%%%%%%%%%%

\subsection{Comparison with the other Fornax clusters}

Cluster 4 differs from the other Fornax clusters in two very important
respects: HB morphology and age.  Because BEA98 used the HB index
(B--R)/(B+V+R) (Lee {\it et al.} 1994) to quantify HB morphologies of
Fornax
clusters 1, 2, 3, and 5, we will also use it here.  However, the
present data are not well suited to detect variables, and the
contamination of the CMD by the field population 
contributes to the uncertainty of this index for cluster 4. 
Nevertheless,
considering that in Fig. 8 all the HB stars are redder than
(V--I)$\simeq$0.63 (i.e. $(V-I)_0\simeq$0.48) with, at most, a handful
of stars bluer than this limit, and considering that the average of
the red-edges of instabilities strips of cluster 1,2,3 and 5 gives
$(V-I)_0$(red edge)=$0.46\pm0.06$, we can safely assume that {\it all}
of the HB stars of cluster 4 are red and therefore estimate
(B--R)/(B+V+R)=--1.0$\pm$0.2.

If we adopt [Fe/H]$\simeq$--2 for cluster 4, as indicated by its RGB,
then it is one of the most extreme examples of a very metal-poor cluster
with
a red HB.  If it is anomalous like R106 (which remains to be
determined), then its [Fe/H] might be as much as 0.4 dex higher.  Even
in this case, it is as an extreme example of the second parameter
effect as the globular clusters Pal 3, Pal 4, Eridanus, and AM-1 (see
Fig. 7 in Lee {\it et al.} 1994), which populate the remote halo of the
Milky Way (as the Fornax dSph does itself).

%%%%%%%%%%%%%%%%%%%%%%%%%%%%%%%%%%%%%%%%%%%%

\subsection{Relative ages}

We will concentrate now on the important issue of the spread in age of
the Fornax clusters, which is most accurately measured among clusters
of very similar composition.  Under the assumption that the Fornax
clusters have the same relative abundances of the elements, the redder
and more gently sloped RGB of cluster 2 indicates it is slightly more
metal-rich than clusters 1, 3, \& 5 ($[Fe/H]=-1.78$ as opposed to
$[Fe/H]\simeq-2$).  We have therefore made separate comparisons of
cluster 4 with clusters 1, 3, \& 5 and with cluster 2.  The following
analysis, which achieves high precision by comparing simultaneously
all the relevant branches of the CMDs (see Buonanno {\it et al.}
1993). A similar procedure was performed by BEA98 for clusters 1, 2, 3
and 5.

In Fig. 12 we show the ridge lines of Fornax clusters 1, 3, 4, \& 5 after
shifting them by the amounts required by their HB luminosities and
reddenings.  The relevant data are reported in Table 4.
Adopting M$_V$(HB)=0.82+0.17[Fe/H], and using the reddenings reported
in Table 4, we obtained the following shifts: $\Delta$V=0.13 for
cluster 1,
$\Delta$V=0.13 for cluster 3,
$\Delta$V=0.40 for cluster 4
and $\Delta$V=0.21  for cluster 5.

>From inspection of Fig. 12 one immediately sees that, in spite of the 
excellent agreement of the RGBs, the TO of
cluster 4 is both brighter and bluer than the others.
This effect clearly deserves further investigation.

The detailed analysis by BEA 98 of the relative ages of clusters
1, 2, 3 and 5 was based on the estimate of two double differential
parameters: $\Delta_V$ and $\delta_{(V-I)}$.  $\Delta_V$ is defined as
$\Delta$$V_{HB}^{TO}${\it (ref)} -- $\Delta$$V_{HB}^{TO}${\it (progr)},
where $\Delta$$V_{HB}^{TO}$ is the difference in luminosity between
the TO point and the HB at the variability strip. The parameter
$\Delta$$_V$ is defined by a pair of clusters, the first being the
reference cluster and the second the current ``program'' cluster.
$\delta_{(V-I)}$ is defined as $\Delta(V-I)_{TO}^{RGB}{\it (ref)}$ --
$\Delta(V-I)_{TO}^{RGB} {\it (progr)}$, where
$\Delta(V-I)_{TO}^{RGB}{\it (ref)}$ is the color difference between
the TO and the base of the RGB and is the equivalent in the V, (V--I)
plane of the $\delta$(B--V) defined by VandenBerg {\it et al.} (1990) in the
V,(B--V) plane.

Under the assumption of similar mixes of elements, the metal
abundances of clusters 1, 3, 4, and 5 are so close that they can be
treated as having the same abundance.  We concentrate on cluster 4
because BEA98 have recently reached the following conclusions
regarding the other clusters:

a) the globular clusters 1, 2, 3 and 5 have essentially the
same age ($\delta t\leq$1 Gyr);

b) the globular clusters 1, 2, 3 and 5 are essentially coeval with the
old,
metal-poor clusters of our Galaxy, M68 and M92;

c) the observed HB morphologies are not explained by differences in
age unless the HB is more sensitive to age differences than has been
previously estimated.  However, a correlation exists between the HB
types and the central densities of the clusters that is qualitatively
similar to one among the globular clusters of the Milky Way.

To obtain a quantitative estimate of the age-differences between
cluster 4 and the other Fornax clusters we use again the procedure
adopted above and the calibrations $\delta$t$_9$=11.6$\Delta$$_V$
(Buonanno {\it et al.} 1993) and $\delta$t$_9$=--107.5$\delta$$_{(V-I)}$
(Buonanno {\it et al.} 1998b) (valid around [Fe/H]=--2 and t=14 Gyr) and
list in Table 4 the differential quantities.
In Table 4, cluster 4 is the reference cluster and the observational
errors have been computed following Buonanno {\it et al.} (1993) and
BEA98. The data for cluster 1, 2, 3 and 5 are from BEA98.

The mean age difference is then: 

$\Delta$t {\it (cl.4 - cl.1,3,5)} = --2.97 $\pm$ 0.54 (from
$\Delta$$_V$)

$\Delta$t {\it (cl.4 - cl.1,3,5)} = --2.83 $\pm$ 0.46 (from
$\delta_{(V-I)}$)

where the associated error is 3 times the standard error, $\sigma$.
>From the mean of the two determinations we conclude that cluster 4 is
about 2.9 Gyr younger than clusters 1, 3, \& 5.

The comparisons made in Fig. 10 between cluster 4 and cluster 2
suggest that cluster 4 is younger than 2 by a significant amount.
Assuming the clusters have similar [Fe/H], $\Delta$t {\it (cl. 4 -
cl. 2)} = --2.5$\pm1.7$ and --2.3$\pm0.6$ , according to $\Delta$$_V$
and $\delta_{(V-I)}$, respectively.

These results for the age differences may change if we relax the
condition that relative abundances of the elements are identical in
the clusters, but without knowing the abundance differences firm
predictions cannot be made.  If cluster 4 is truly analogous to R106,
it may be more Fe rich but also more [$\alpha$/Fe] poor than the other
Fornax clusters.  Because these differences have offsetting effects on
both $\Delta$$V_{HB}^{TO}$ and $\Delta(V-I)_{TO}^{RGB}$, the relative ages
of the clusters may not be affected by much.  There is strong evidence,
therefore, for a several Gyr range in age among the Fornax clusters.

It is important to see if this age difference is consistent with
the very red HB morphology of cluster 4.  Since cluster density may
affect HB
morphology (Buonanno {\it et al.} 1997, BEA 1998), this comparison is
best done between clusters 4 and 3 which have nearly identical central
densities (Webbink 1985).  These clusters differ by 1.5$\pm$0.2 in HB
type.  Nearly the same differences exist between cluster 4 and 
clusters 5 and 2, whose HBs are only slightly redder than that of cluster 3.

The synthetic HB calculations by Lee {\it et al.} (1994) show that for
[Fe/H]=--2 and an absolute age near 14 Gyr, an age difference of
$\delta$t$\simeq$3 Gyr is expected to produce a difference in HB type
of about this size.  This can be illustrated using the comparison that
Lee {\it et al.} (1994) made between the two Milky Way globular
clusters, R106 and NGC 6397, whose HB types (--0.82 and 0.93,
respectively) differ by 1.75.  From the analysis of their Fig. 16,
Lee {\it et al.} (1994) concluded that for a the metallicity near
[Fe/H]=--1.9 and for an absolute age of 15 Gyr, this difference in HB
type would be explained if R106 is younger than NGC6397 by 3.6 Gyr
(under the assumption of variable mass loss on the RGB).  This age
difference is in agreement with the result of Buonanno {\it et al.} (1993)
who, from the TO luminosities, concluded that R106 is about 4 Gyr
younger than typical metal-poor clusters such as NGC6397.  The age
difference predicted by Lee {\it et al.} (1994) under the assumption
of constant mass loss is somewhat too large, 5.4 Gyr, which
illustrates the uncertainty of any estimate of age differences from HB
morphology.  It is important to add that the unusual mix of elements
in R106 is unlikely to be the major reason why its HB is so red in
comparison to NGC6397 and most other metal-poor globular clusters.  As
we have discussed previously, the measurements of Brown {\it et al.} (1997)
indicate that R106 has a larger [Fe/H] than NGC6397 but probably also
a smaller [$\alpha$/Fe], which have opposite effects on HB morphology.
While the uncertainties are many, the difference in age between
cluster 4 and the other Fornax clusters is of the correct sign and
magnitude to explain their difference in HB type.

\subsection{Comparison with the Milky Way cluster Ruprecht 106}

As noted in the Introduction, the tidal destruction of dwarf satellite
galaxies has been widely discussed as a possible origin for the outer
halo of the Milky Way, in part because this may explain the greater
dispersion in HB types, particularly the relatively high frequency of
red HB types, among the metal-poor outer halo (R$_{gc}$$\geq$8 kpc)
clusters (e.g., Searle \& Zinn 1978).  This possibility has motivated
several comparisons between the Fornax clusters and more recently the
clusters of the Sagittarius dSph with the globular clusters of the
outer halo (Zinn 1993; Smith {\it et al.} 1998; Marconi {\it et al.}
1998).  The close similarity between cluster 4 and R106, which we now
discuss, lends weight to the hypothesis that R106 and the other
unusual outer halo clusters were once members of dwarf galaxies.

The comparison of the fiducial line of cluster 4 with that of R106
(Buonanno {\it et al.} 1993) is made in Fig. 13.
The line for Fornax cluster 4 has been dereddened and
shifted according to the quantities already adopted for Fig. 12. The
mean points of R106 have been shifted by E(V--I)=0.27 and
$\Delta$V=+3.28.  The first quantity is nearly identical to the
reddening found by Buonanno {\it et al.} (1993), who measured
E(V--I)=0.23, and the second is exactly the difference between the
absorption-free HB magnitudes (V$_{HB}$(R106)=17.85).
To estimate the age difference between cluster 4 and R106, we use again
the
procedure adopted above. From Fig. 13 we find 
$\Delta _V(cl.4-R106)=0.02$,
$\delta_{(V-I)}(cl.4-R106)=-0.05$,
and then $\delta t$ = 0.2 Gyr and $\delta t$=0.5 Gyr,
respectively.  Cluster 4 and R106 are therefore essentially coeval.
While R106 and cluster 4 appear to be very similar as far as age, HB
morphology and the metallicity are concerned, they are very different
in central density (log $\rho$$_0$(cl. 4)=3.936,
log$\rho$$_0$(R106)=1.216 M$_{\odot}$pc$^{-3}$).  This difference
appears to have had little if any effect on the HB morphologies of the
clusters.

The investigations of the Fornax clusters reported here and in BEA98
have not found compelling evidence that the second parameter can be
identified with only one quantity.  Our results suggest that age
differences alone may explain the difference in HB morphology between
cluster 4 and the other clusters, if one accepts the calculations of
Lee {\it et al.} (1994) for the variable mass-loss case.  However,
these same calculations predict a larger age difference than is
observed between cluster 1 and clusters 2, 3, and 5 (BEA98).  This
suggests that the second parameter phenomenon may be caused by a
mixture of age with other effects.

As discussed earlier by Buonanno {\it et al.} (1997) and BEA98, it
is attractive to identify cluster density as an additional factor,
which presumably affects HB morphology by influencing the amount of
mass loss on the RGB.  The effective temperatures of HB stars depend
on their envelope masses, and this sensitivity is greatest among the
blue HB stars, which have the lowest envelope masses.  If in clusters
of high density the stars evolving on the RGB loose a little
additional mass through stellar encounters, then the effect on HB
morphology will be greatest in clusters that have blue HBs for another
reason, such as very old age.  This may explain why density
appears to be correlated with the HB morphology of blue HB clusters
and with the oddities of blue HBs, such as ``blue tails'', but at the
same time appears to have little effect on the morphologies of red HB
clusters such as cluster 4 and R106.  Theoretical investigations of
this question, the more general question of the sensitivity of HB
morphology to age and other 
factors (rotation, mixing etc.), and the mass-loss mechanism(s)
are urgently needed.

%%%%%%%%%%%%%%%%%%%%%%%%%%%%%%%%%%%%%%%%%%%%%%%%%%%%%%%%%%%%%%%%%%%%%%%%%%%%

\section{Summary and conclusions}

We have constructed deep CMDs of cluster 4 in Fornax and of the field
population around the cluster using data from the HST archive.  From
these CMDs, we measured the metal abundances and estimated the ages of
their stellar populations.

Our results for the field population of Fornax are in good agreement
with previous investigations in that they reveal a very long period of
star formation ($\simeq$12 to 0.5 Gyr).  While a small fraction of the
field stars may be coeval with the Fornax clusters and as metal poor,
the majority of them are significantly younger than the youngest
cluster and more metal rich.  The multiple SGBs in the field CMD
suggest that the rate of star formation was not constant but resembled
more bursts.

In contrast to the field population, all 5 globular clusters in Fornax
appear to be older than about 10 Gyrs, but not, however, without a
significant spread in age.  BEA98 showed that clusters
1, 2, 3, and 5 are essentially coeval ($\delta t\leq$1 Gyr) with each
other and with the very metal-poor globular clusters in the Milky Way.
Our results indicate that cluster 4 is younger,
$\delta t\simeq$3 Gyr, than the other Fornax clusters.  The RGB of
cluster 4
is very steep, which under the standard assumptions is a sign that it
is very metal-poor, [Fe/H]$\simeq$--2.  The very close similarity
between the CMDs of cluster 4 and R106 suggests, however, that it may
be also like R106 in having an unusual mix of elements, a
possibility that warrants further investigation.  

Although the uncertainties are large, the very red HB of cluster 4 may
be explained entirely by its relatively young age.  In contrast, BEA98
found that age differences alone were unlikely to account for the
range in HB types among the other Fornax clusters, and they suggested
that cluster density also played a role.  Our comparison between
cluster 4 and R106, which are very similar in CMD morphology despite
very different central densities, suggests that cluster density has at
most a small effect among relatively youthful clusters that have very
red HBs.

The two most massive dSph galaxies orbiting the Milky Way, Fornax and
Sagittarius, have their own globular cluster systems in which there
are different cluster-to-cluster variations in metal abundance, HB
type, and age (see also Marconi {\it et al.} 1998; BEA98).  The similarities
found here and in BEA98 between the Fornax clusters and both
``normal'', e.g. M68 and M92, and ``anomalous'', e.g. R106, halo
clusters supports the hypothesis that the tidal destruction of similar
galaxies in the past, as is now happening to the Sagittarius dSph, is
the reason for the diversity in properties among the outer halo
globular clusters.

%%%%%%%%%%%%%%%%%%%%%%%%%%%%%%%

\begin{acknowledgements}

This research is based on observations with the NASA/ESA {\it Hubble
Space Telescope} obtained at the Space Telescope Science Institute,
which is operated by Association of Universities for Research in
Astronomy, Inc., under NASA contract NAS 5-26555.  The support of the
CNAA for M.C. and the NSF (AST-9319229 \& AST-9803071) and STScI
(GO-05917.01-94A) for R.Z. is gratefully acknowledged.

We acknowledge J. R. Westphal, P.I. of the GTO proposal WFC 5637 for designing 
the very useful observations we used in this paper. 
We also thank A. Chieffi for providing us updated theoretical isochrones and 
for helpful suggestions and comments. 

\end{acknowledgements}

\pagebreak

%%%%%%%%%%%%%%%%%%%%%%%%%%%

\newpage

% ----------------------------------------------------------

\centerline{\bf Figure captions}

\figcaption[figure1.eps]{Fornax cluster 4 (WFPC2 mosaic).
Objects enclosed in numbered squares are selected bright stars listed in 
Table 1a for identification purpouse.}

\figcaption[figure2.eps]{The CMD obtained for the three WFPC2 fields. 
Photometric errors, for both magnitudes and colors, are shown on the
left of the diagram.}

\figcaption[figure3.eps]{The stellar density profile of our field. Errors
are showed as vertical bars at data points.}

\figcaption[figure4.eps]{The CMD of the {\it field} of Fornax.}

\figcaption[figure5.eps]{The ridge line of the Fornax field and of the GC M5.}

\figcaption[figure6.eps]{CMD of the field, compared with the ridge line 
of M5 and Yale theoretical isochrones for selected ages.}

\figcaption[figure7.eps]{As for Fig. 6, but with the isochrones of 
Chieffi, Straniero \& Limongi (private communications) for the same ages.}

\figcaption[figure8.eps]{CMD of Fornax cluster 4.}

\figcaption[figure9.eps]{The ridge line of Fornax cluster 4
compared with ridge lines of GGCs in DCA90. Left to
right, M15 ([Fe/H]=--2.17), NGC6397 (--1.91), M2 (--1.58),
NGC6752 (--1.54), NGC1851 (--1.29) and 47 Tuc (--0.71). Metallicities
are taken also from DCA90.}

\figcaption[figure10.eps]{The ridge lines of clusters 2 and 4 under two 
different assumptions for the reddening of cluster 4}

\figcaption[figure11.eps]{The ridge lines of cluster 4 and of M68.}

\figcaption[figure12.eps]{The fiducial lines of Fornax clusters 1,3,4,5.}

\figcaption[figure13.eps]{The fiducial lines of cluster 4 and of
Ruprecht 106.}


\begin{thebibliography}{}

\bibitem{}
Azzopardi, M., 1994, in "The Local Group: Comparative and Global
	Properties", edited by A. Layden, R.C. Smith \& J. Storm),
	129, ESO Conference Series

\bibitem{}
Beauchamp, D., Hardy, E., Suntzeff N.B. \& Zinn, R., 1995, AJ, 109, 1629

\bibitem{} 
Bertelli, G., Bressan, A., Chiosi, C., Fagotto, F. \& Nasi,
E. 1994, A \&AS, 106, 275

\bibitem{}
Brown, J.A., Wallerstein, G. \& Zucker, D. 1997, AJ, 114, 180

\bibitem{}
Buonanno, R., 1988, MIDAS Manual, Chapter 5

\bibitem{}
Buonanno, R., Corsi, C.E., Bellazzini, M., Ferraro, F. \& Fusi Pecci,
F.,
	1997, AJ, 113, 706

\bibitem{}
Buonanno, R., Corsi, C.E., Fusi Pecci, F., Richer, H. \& Fahlman, G.C.,
	1993, AJ, 105, 184

\bibitem{}
Buonanno, R., Corsi, C.E., Fusi Pecci, F., Hardy, E. \& Zinn R., 
	1985, A\&A, 152, 65 (BEA85)

\bibitem{}
Buonanno, R., Corsi, C.E., Zinn, R., Fusi Pecci, F., Hardy, E. Suntzeff,
N.B., 1998a, ApJL, 501, 33 (BEA98)

\bibitem{}
Buonanno, R., Corsi, C.E. \& Fusi Pecci, F., 1989, A\&A, 216, 80

\bibitem{}
Buonanno, R., Corsi, C.E., Pulone, L., Fusi Pecci, F. \& Bellazzini, M.,
	1998b, A\&A, 333, 505

\bibitem{}
Cardelli, J.A., Clayton, G.C. \& Mathis, J.S., 1989, ApJS, 345, 245

\bibitem{}
Carretta, E. \& Gratton, R.G. 1997, A\&AS, 121, 95

\bibitem{}
Chaboyer, B., Demarque, P., Kernan, P.J. \& Krauss, L.M. 1998, ApJ, 494,
96

\bibitem{ }
Cool, A.M., King, I.R. 1995, in ``Calibrating HST: Post
  Servicing Mission'', eds. A. Koratkar \& C. Leitherer (Baltimore:
  STScI), p. 290
     
\bibitem{}
Da Costa, G.S. 1998, in ``Stellar Astrophysics for the
	Local Group'', eds. A. Aparicio, A. Herrero, \& F. Sanchez, Cambridge:
	Cambridge Univ. Press, p. 351

\bibitem{}
Da Costa, G.S. \& Armandroff, T.E., 1990, AJ, 100, 162 (DCA90)

\bibitem{}
Demarque, P., Chaboyer, B., Guenther, D., Pinsonneault, M., 
	Pinsonneault, L., \& Yi, S. 1996, 
	http://shemesh.gsfc.nasa.gov/iso.html

\bibitem{}
Dubath P., Meylan G. \& Mayor M., 1992, ApJ, 400, 510

\bibitem{}
Francois, P., Danziger, I.J., Buonanno, R., Perrin, M.,
	1997, A\&A, 327, 121

\bibitem{}
Fusi Pecci, F., Bellazzini, M., Cacciari, C. \& Ferraro,
	F.R. 1995, AJ, 110, 1664

\bibitem{} 
Hardy, E., Buonanno, R., Corsi, C.E., Janes, K.A., \&
	Schommer, R.A. 1984, ApJ, 278, 592

\bibitem{}
Harris, H.C. \& Canterna R., 1977, AJ, 82, 798

\bibitem{}
Holtzman, J.A., Burrows, C.J., Casertano, S., Hester, J. {\it et al.},
	1995, PASP, 107, 1065

\bibitem{}
Johnson, J.A. \& Bolte, M., 1998, AJ, 115, 693	

\bibitem{}
Lee, Y-W., Demarque, P. \& Zinn, 1990, ApJ, 350, 155

\bibitem{}
Lee, Y-W., Demarque, P. \& Zinn, R., 1994, ApJ, 423, 248

\bibitem{}
Lee, M.G., Freedman, W.L. \& Madore, B., 1993, ApJ, 417, 553

\bibitem{}
Marconi, G., Buonanno, R., Castellani, M., Iannicola, G.,
	Molaro, P., Pasquini, L. \& Pulone, L.,
	1998, A\&A, 330, 453		

\bibitem{}
Mateo, M. 1998, An. Rev. Astr. Astrophys., 36, 435

\bibitem{}
Mighell, K.J. 1997, AJ, 114, 1458

\bibitem{}
Popowski, P. \& Gould, A. 1998, ApJ, 506, 271

\bibitem{}
Salaris, M., Chieffi, A., \& Straniero, O. 1993, ApJ, 414, 580

\bibitem{}
Sarajedini, A. 1994, AJ, 107, 618

\bibitem{}
Sarajedini, A. \& Layden, A. 1997, AJ, 113, 264

\bibitem{}
Sarajedini, A., Chaboyer, B., Demarque, P., 1997, PASP, 109, 1321

\bibitem{}
Searle, L. \& Zinn, R. 1978, ApJ, 225, 357

\bibitem{} 
Smecker-Hane, T., Stetson, P.B., Hesser, J.E., \& Lehnert,
M.D. 1994, AJ, 108, 507

\bibitem{}
Smecker-Hane, T., Stetson, P.B., Hesser, J.E. \& VandenBerg, D. A.
1996, ASP Conf. Ser. 98, p. 328

\bibitem{}
Smith, E.O., Neill, J.D., Mighell, K.J., \& Rich, R.M. 1996, AJ, 111,
1596

\bibitem{}
Smith, E.O., Rich, R.M., \& Neill, J.D. 1997, AJ, 114, 1471

\bibitem{}
Smith, E.O., Rich, R.M., \& Neill, J.D. 1998, AJ, 115, 2369

\bibitem{}
Stetson, P.B., 1987, PASP, 99, 191

%\bibitem{}
%Stetson, P.B. 1997, Baltic Astron., 6, 3

\bibitem{}
Stetson, P.B., Hesser, J.E., \& Smecker-Hane, T. A. 1998, PASP 110, 533

\bibitem{}
Straniero, O. \& Chieffi, A., 1991, ApJS, 76, 525

\bibitem{}
Vandenberg, D.A., Bolte M. \& Stetson, P.B., 1990, AJ, 100, 445

\bibitem{}
Vandenberg, D.A., Stetson, P.B. \& Bolte, M., 1996, ARA\&A, 34, 461

\bibitem{}
Walker, A.R., 1994, AJ, 108, 555

\bibitem{}
Webbink, R.F., 1985, in "Dynamics of Star Cluster", edited by
	J. Goodman \& P. Hut, IAU Symphosium 113, p. 541

\bibitem{}
Welch, D.L., McLaren, R.A., Madore, B.F. \& McAlary, C.W., 
	1987, ApJ, 321, 162

\bibitem{} 
Zinn, R. 1993, in ``The Globular Cluster-Galaxy
Connection'', A.S.P. Conf. Series vol. 48, 302

\bibitem{}
Zinn, R. \& Persson, S.E., 1981, ApJ, 247, 849

\bibitem{}
Zinn, R. \& West, M., 1984, ApJS, 55, 45	


\end{thebibliography}
\end{document}